\documentclass[a4paper,11pt]{article}

\usepackage{contribution}

\newcommand{\weblink}[2][]{%
    \ifthenelse{\equal{#1}{}}%
    {\textnormal{\url{#2}}}%
    {\textnormal{\href{#2}{#1}}}%
}

\newcommand{\acknowledgements}[1]{%
  \bigskip\bigskip
  \textsf{\textbf{\Large Acknowledgements}} \\[2ex]
  {#1}
  \bigskip
}

\def\beq{\begin{equation}}
\def\eeq#1{\label{#1}\end{equation}}
\def\eeqn{\end{equation}}

\def\beqa{\begin{eqnarray}}
\def\eeqa#1{\label{#1}\end{eqnarray}}
\def\eeqan{\end{eqnarray}}

\let\bar=\overbar

\def\Dslash{\not{\hbox{\kern-4pt $D$}}}
\def\dslash{\not{\hbox{\kern-2pt $\del$}}}

\def\msb{{\bar{\ssstyle M \kern -1pt S}}}

\newcommand{\contribution}[7][]{%
  \clearpage
  \thispagestyle{plain}
  \ifthenelse{\equal{#1}{}}
  {\hypersetup{pdftitle={#2}}}
  {\hypersetup{pdftitle={#1}}}
  \hypersetup{pdfauthor={{#3} {#4}}}
  {\centering\normalfont\LARGE\bfseries\sffamily #2 \par\nobreak}
  \lhead{}
  \chead{%
    \textit{\footnotesize XIV International Conference on Hadron Spectroscopy
      (\weblink[\textit{hadron2011}]{http://www.hadron2011.de}), 13-17 June 2011, Munich, Germany}%
  }
  \rhead{}
  \bigskip
  \begin{center}
    {#3} {#4}\ifthenelse{\equal{#6}{}}{}{\footnote{\weblink[#6]{mailto:#6}}}
    \ifthenelse{\equal{#7}{}}{}{#7} \\
    \textit{#5}
  \end{center}
  \bigskip
}

\renewcommand{\abstract}[1]{%
  \begin{center}
    \begin{minipage}{0.85\textwidth}
      \begin{footnotesize}
        #1
      \end{footnotesize}
    \end{minipage}
  \end{center}
  \bigskip
}

\begin{document}

%
%
%
%
%
{  

\makeatletter
\@ifundefined{c@affiliation}%
{\newcounter{affiliation}}{}%
\makeatother
\newcommand{\affiliation}[2][]{\setcounter{affiliation}{#2}%
  \ensuremath{{^{\alph{affiliation}}}\text{#1}}}
%

\contribution[Pivotal year for GPDs]  
{A pivotal year for \\ Generalized Parton Distributions}  
{Herv\'e}{Moutarde}  
{\affiliation[IRFU/Service de Physique Nucléaire, CEA, Centre de Saclay, F-91191 Gif-sur-Yvette, FRANCE]{1}} 
{herve.moutarde@cea.fr}  
{\!\!$^,\affiliation{1}$, J.~Ball\affiliation{1}, G.~Charles\affiliation{1}, B.~Moreno\affiliation{1}, F.~Sabatié\affiliation{1}, and S.~Procureur\affiliation{1}}
%

\abstract{%
The field of Generalized Parton Distribution (GPD) benefited from a wealth of exclusive reactions measurements since 2000. Extraction of GPDs from observables has begun with this first generation of experiments. In the short and mid-term future the programs of COMPASS-II and JLab at 12~GeV will enlarge the existing kinematic domain and open an era of high-precision measurements. Tools for phenomenological analysis are being developed at the same time and will be mature to handle these forthcoming measurements. In the long-term future the EIC project will allow an ambitious spin program in the small to intermediate $x_B$ domain.
}
%

\section{Introduction}

The Deeply Virtual Compton Scattering (DVCS) process was early recognized as the cleanest way to access Generalised Parton Distributions (GPD) and has been so far a very active field of research. However a flavour separation of GPDs will require Deeply Virtual Meson Production (DVMP) measurements (see reviews~\cite{Goeke:2001tz, Diehl:2003ny, Belitsky:2005qn, Boffi:2007yc} and references therein). 

One of the goals of this program is imaging the nucleon in 3d. Present observables may already be fit, giving indications on the actual sensitivity of observables to GPDs and hints to elaborate robust extracting methods.

The first section is a brief reminder. The second part outlines some results of the extraction of GPDs as of 2011. The last section mentions promising future prospects.

%


\section{DVCS at leading twist and leading order}

Four GPDs $H$, $E$, $\tilde{H}$ and $\tilde{E}$ describe DVCS at leading twist and leading order, but the cross section depend on the Compton Form Factors (CFF) $\mathcal{H}$, $\mathcal{E}$, $\tilde{\mathcal{H}}$ and $\tilde{\mathcal{E}}$ (see ref.~\cite{Belitsky:2001ns}). The GPDs depend on the generalised Bjorken variable  \cite{Belitsky:2001ns} $\xi$ (or equivalently on the standard Bjorken variable $x_B$), the virtuality of the initial photon $Q^2$ and the square momentum transfer between initial and final protons $t$. The CFFs are convolutions along the variable $x$ of GPDs with known kernels. This integration kernel yields a real and an imaginary part to a CFF. They are related by fixed-$t$ dispersion relations \cite{Anikin:2007yh, Diehl:2007jb} involving a subtraction constant related to the $D$-term \cite{Polyakov:1999gs}. However the $D$-term is poorly known and most of DVCS measurements were made in the region $\xi \leq 0.5$. Using dispersion relations thus rely on models and may introduce biases in the extraction of GPDs from DVCS data. For that reason the real and imaginary parts of CFFs are taken as independent in some fitting procedures.

%


\section{Status of GPD analysis}

\subsection{Counting degrees of freedom}

The problem consists in extracting four functions $H$, $E$, $\tilde{H}$ and $\tilde{E}$ of three variables\footnote{We omit the $Q^2$-dependence which is governed by the QCD evolution equations.} $x$, $\xi$ and $t$ for each quark flavour ($u$, $d$ and $s$). The knowledge of these functions should provide us Parton Distribution Functions (PDF) and Form Factors (FF). A naive counting quickly shows that building a flexible yet robust GPD parametrization is very involved because the number of free parameters has to remain low to keep the fit tractable and the numerics under control.

\subsection{Local fits}

This approach is detailed in ref.~\cite{Guidal:2008ie} and assumes the independence of the real and imaginary parts of CFFs. The mains assumptions are the validity of the twist-2 analysis of existing DVCS measurements and a negligible contribution of $\mathop{\mathrm{Im}}\tilde{\mathcal{E}}$. Each kinematic bin $( x_B, t, Q^2 )$ is taken independently of the others, and the seven values $\mathop{\mathrm{Re}}\mathcal{H}$, $\mathop{\mathrm{Im}}\mathcal{H}$, $\mathop{\mathrm{Re}}\mathcal{E}$, $\mathop{\mathrm{Im}}\mathcal{E}$, $\mathop{\mathrm{Re}}\tilde{\mathcal{H}}$, $\mathop{\mathrm{Im}}\tilde{\mathcal{H}}$ and $\mathop{\mathrm{Re}}\tilde{\mathcal{E}}$ are extracted simultaneously. Nothing prevents large fluctuations between two neighbouring kinematic bins. In the following we will refer to these fits as \emph{local fits}. 

The model-dependence is almost as low as possible but the problem is often ill-posed because the harmonic structure of a given observable (\textit{e.g.} a BSA) at fixed kinematic usually depends on less than 7 parameters. Thus it is mandatory to combine different observables on the same kinematic bins (for example BSAs and Target Spin Asymmetries (TSA) or Beam Charge asymmetries (BCA)) and to assume boundaries for the values the CFFs may take.

\subsection{Global fits}

In the spirit of the work done on PDFs and FFs, \emph{global fits} require a physically motivated parametrization of GPDs and deal with all observables on all kinematic bins at once. 

The free coefficients entering the expressions for GPDs are determined either from PDFs and FFs or from DVCS data. Two such studies have been reported so far for DVCS. In the first case \cite{Kumericki:2009uq} all DVCS measurements on an unpolarized target are included in the fit in a (mostly) $H$-dominant framework. Fixed-$t$ dispersion relations are used as a key ingredient. In the second case \cite{Goldstein:2010gu} CLAS and HERMES measurements are used in the fit, and JLab Hall A data are computed as a prediction of the fitting Ansatz~; all four GPDs $H$, $E$, $\tilde{H}$ and $\tilde{E}$ are described. 

The extrapolation outside the physical region (for example the interesting limits $t \rightarrow 0$ or $\xi \rightarrow 0$) is possible but an assessment of the parametrization uncertainties is a complex task in itself.

\subsection{Hydrid~: Local~/~global fits}

This \emph{hydrid} fitting procedure \cite{Moutarde:2009fg} is a combination of the previous two methods and has been applied with the main assumption of $H$-dominance and twist-2 accuracy. It involves a parametrization with the minimal properties~: smoothness, leading order $Q^2$ evolution and polynomial parametrization. Since the functional is otherwise arbitrary, it is \textit{a posteriori} validated by the quality of the fit.

The model dependence is tested by a systematic comparison to local fits and an estimate of the systematic error induced by the $H$-dominance hypothesis. The good agreement of the two methods (local fits vs global fits) is a necessary consistency check of this approach. The treatment of systematic errors is however a weak point since there is no rigorous procedure to evaluate them and only estimates are given using joint models and fits.

\subsection{Neural networks}

\emph{Neural network} fits had been successfully performed for PDFs but their use for GPD extraction is quite recent. First results are described in \cite{Kumericki:2011rz} within the $H$-dominance assumption. It is worth noting that it is a new development in the field of GPD extraction. It is hoped it will allow better estimates of errors when extrapolating outside the fitting domain.

\subsection{Key results}

To our knowledge, the first extraction of a combination of GPDs was made in \cite{MunozCamacho:2006hx} and the progress have been continuous so far. The different methods all agree on the following key results~:
\begin{itemize}
\item Dominance of twist 2 and validity of a GPD analysis of DVCS data. This is no trivial result considering the low $Q^2$ range accessible in Hermes and JLab kinematics.
\item $\mathop{\mathrm{Im}}\mathcal{H}$ is the best determined quantity with an overall fair agreement between all estimates. However large uncertainties subsist on $\mathop{\mathrm{Re}}\mathcal{H}$.
\item There are already some indications about the invalidity of the $H$-dominance hypothesis even with the experimental statistical accuracy achieved so far.
\end{itemize}

\subsection{A trivial universality check}

Since GPDs have been modelled or extracted from DVCS and DVMP measurements, it is natural to check if these objects are identical. Although far from a satisfactory treatment of the universality question, as a first step we can take GPDs obtained in DVMP studies, predict DVCS observables and compare them to existing measurements. The predictions of the Kroll~-~Goloskokov model (see \cite{Goloskokov:2009ia} and references therein) are depicted in Fig.~\ref{FigBSA} (left). In spite of the fair agreement between this model and the extracted values of $\mathcal{H}$, the resulting DVCS prediction tends to overshoot the data. This is reminiscent of the current version of the VGG GPD model \cite{Goeke:2001tz}.

\begin{figure}[htb]
	\begin{tabular}{cc}
		\includegraphics[width=0.5\textwidth]{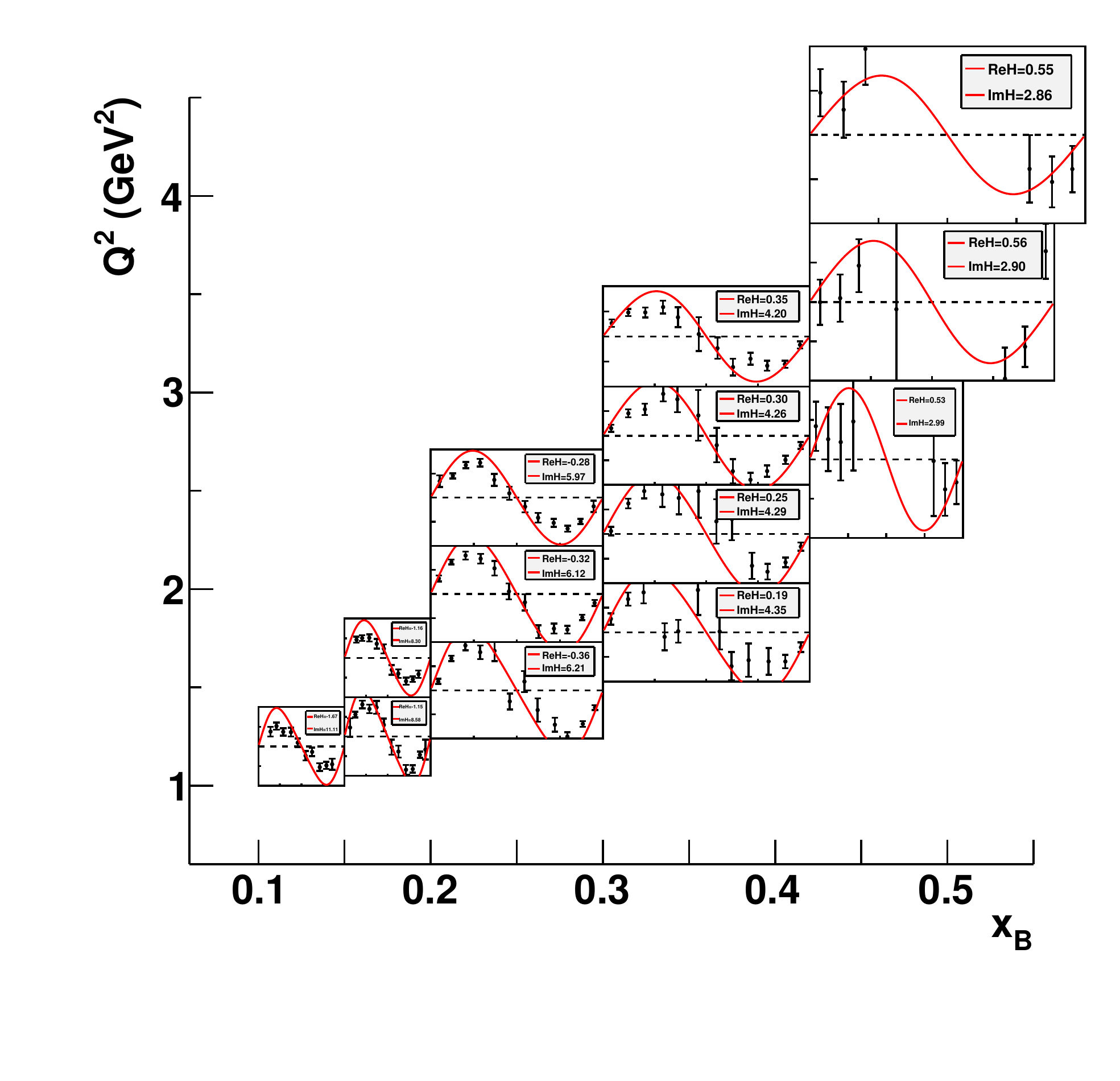}
		&
		\includegraphics[width=0.5\textwidth]{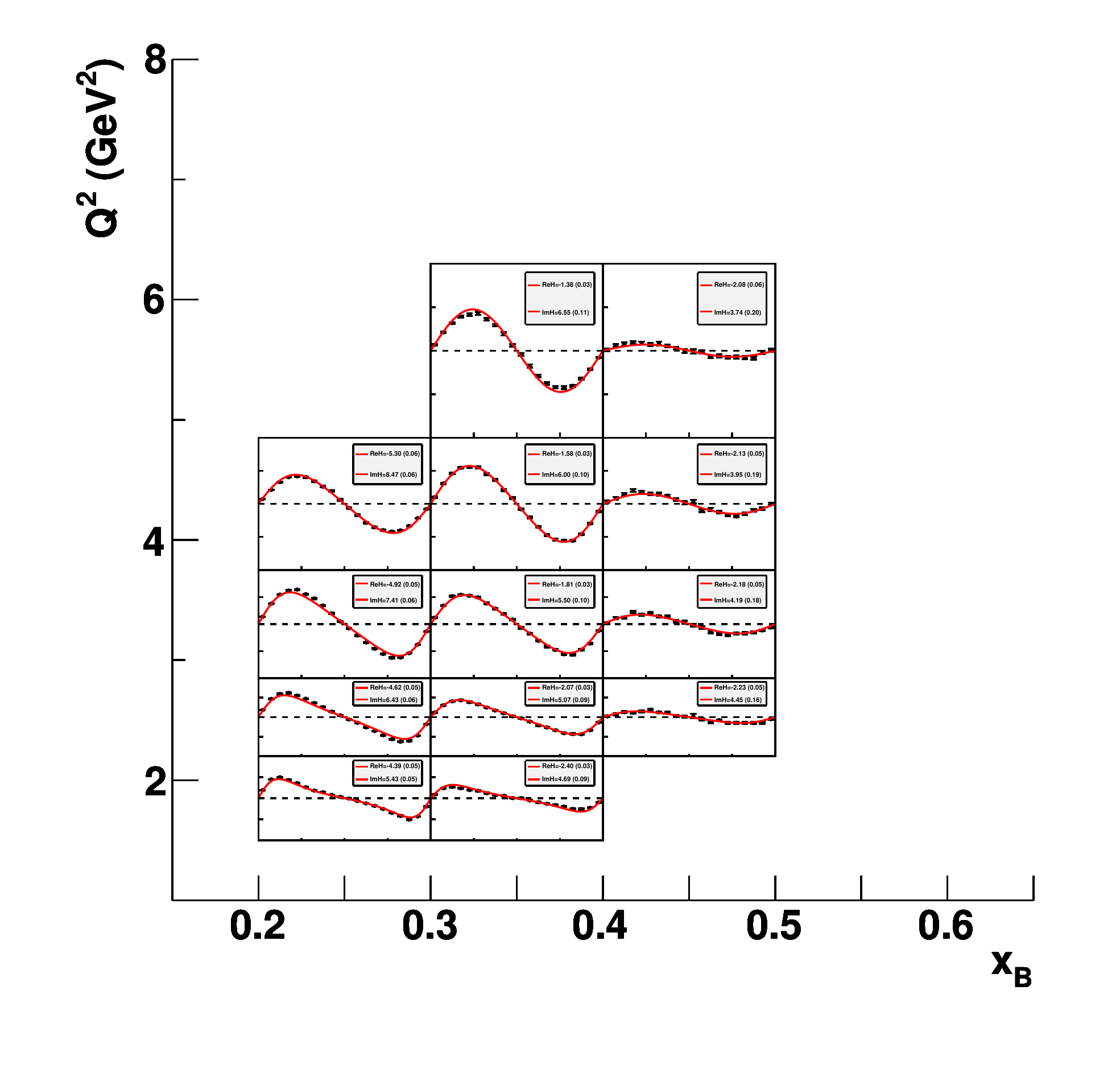}
	\end{tabular}
	\caption{Measured CLAS BSAs for  $-0.4 < t < -0.2~\textrm{GeV}^2$. The red lines correspond to the prediction of the Kroll~-~Goloskokov GPD model (left). Projected CLAS12 BSAs for  $-0.35 < t < -0.25~\textrm{GeV}^2$. Statistical errors are smaller than data point bullets. The red lines correspond to a fit of the data. (right)}
	\label{FigBSA}
\end{figure} 

%


\section{Prospects}

\subsection{Facilities and projects}

In 2012 COMPASS-II will start running its GPD program, giving access to several observables with beam spin and charge differences, in between the kinematic domain of collider and fixed-target experiments.

By 2015 JLab's 12~GeV upgrade will reach a time of high-precision measurements with an expected few percent accuracy. See Fig.~\ref{FigBSA} (right) for projected data and a tentative fit. Despite an apparent agreement, the high accuracy of the data does not allow a fit assuming $H$-dominance any more.

The science case for an electron-ion collider (EIC) has been developed since 2010 \cite{Boer:2011fh}. In the 2020 decade we expect to produce a wealth of GPD related spin observables (see projected data in Fig.~\ref{FigBSAEIC}) with the aim of imaging the sea quark and gluon content of the nucleon.

\begin{figure}[htb]
	\begin{center}
		\includegraphics[width=0.5\textwidth]{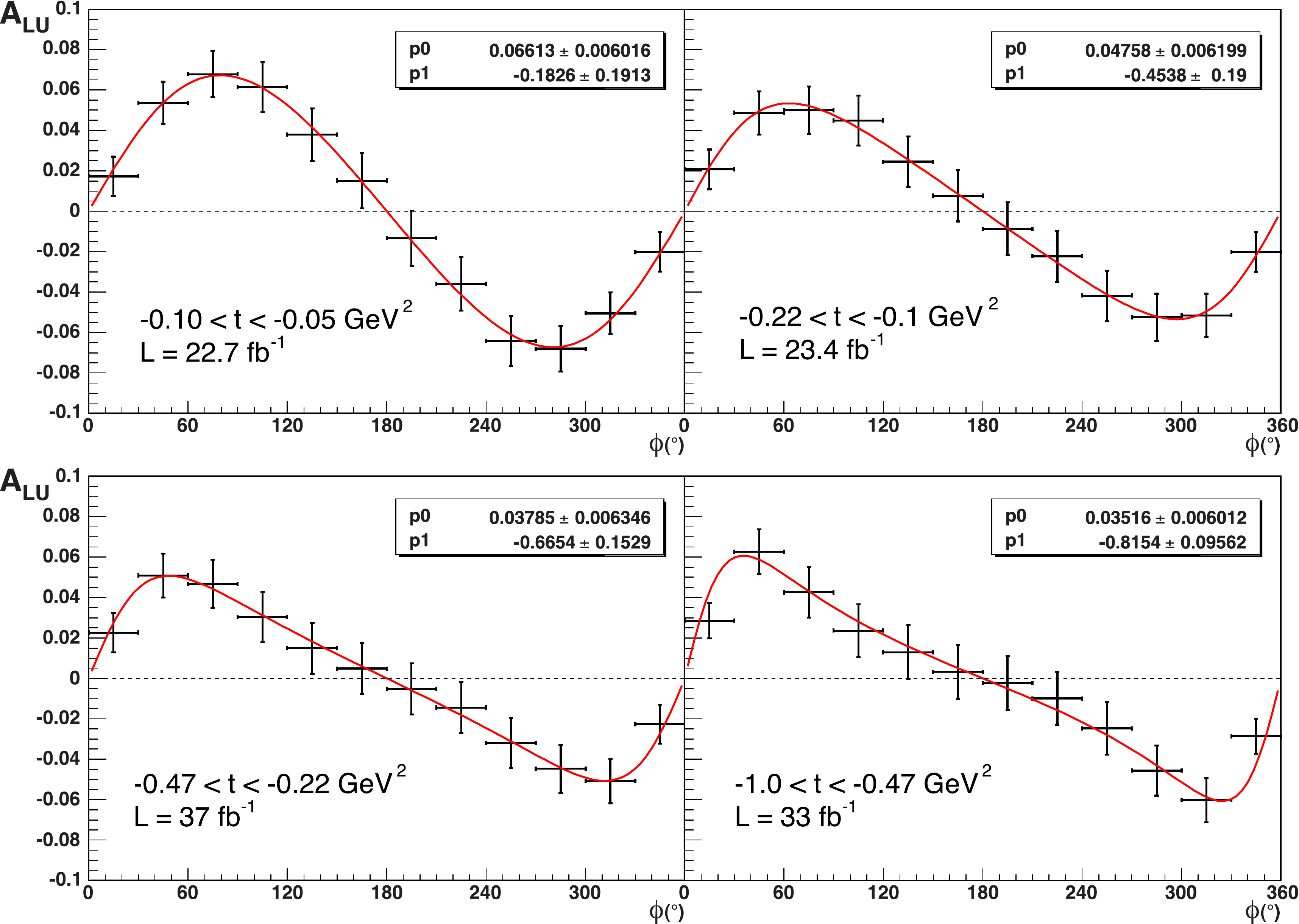}
		\caption{Photon electroproduction beam spin asymmetries for the $20\times 250$ EIC configuration, in the typical kinematic bin: $1.58 \cdot 10^{-3} < x_B < 2.51 \cdot 10^{-3}$, $3.16 < Q^2 < 5.61~\rm{GeV}^2$ for four different $t$-bins as shown on each plot. Up to about 3 months of beam time with 50\% efficiency is necessary to achieve reasonable statistics in some of the bins.}
		\label{FigBSAEIC}
	\end{center}
\end{figure}

\subsection{Phenomenological tools for GPD analysis}

A software platform is being developed to handle the phenomenological analysis of past and forthcoming data. It contains a database of experimental results, a database of theoretical predictions, a fitting engine and a visualizing software to compare measurements and model expectations. Modularity is a key-issue of the design~: the same libraries are used to extract GPDs from measurements and to design new experiments connected to experimental descriptions and event generators.

\section{Conclusions}

During the 2000s GPD studies have made tremendous progress, with important measurements and encouraging first results on the extraction of GPDs. However a robust and efficient fitting strategy for DVCS and DVMP is still needed.

The 2010s will be very exciting times for GPDs~: new facilities, high-precision measurements and bigger kinematic domains will constrain the phenomenology of GPDs. To this end a platform dedicated to global GPD analysis is currently under development.

\acknowledgements{%
The author would like to thank S.~Goloskokov, P.~Guichon, P.~Kroll, K.~Kumeri\v{c}ki, D.~Müller and K.~Passek-Kumeri\v{c}ki for many fruitful and stimulating discussions. The author also thanks the organizers of the 14th International Conference on Hadron Spectroscopy held in Munich (June 13 -June 17, 2011). This work was supported in part by the Commissariat à l'Energie Atomique and the GDR n° 3034 Physique du Nucleon.
}


%

}  



\begin{thebibliography}{99}
  
\bibitem{Goeke:2001tz} 
   K.~Goeke, M.V.~Polyakov and M.~Vanderhaeghen, Prog. Part. Nucl. Phys. {\bf 47}, 401 (2001).
\bibitem{Diehl:2003ny} 
   M.~Diehl, Phys. Rept. {\bf 388}, 41 (2003).
\bibitem{Belitsky:2005qn} 
   A.V.~Belitsky and A.V.~Radyushkin, Phys. Rept. {\bf 418}, 1 (2005).
\bibitem{Boffi:2007yc} 
   S.~Boffi and B.~Pasquini, Riv. Nuovo Cim. {\bf 30}, 387 (2007).
\bibitem{Belitsky:2001ns} 
   A.V.~Belitsky, D.~Mueller and A.~Kirchner, Nucl. Phys. B {\bf 629}, 323 (2002).
\bibitem{Anikin:2007yh} 
   I.V.~Anikin and O.V.~Teryaev, Phys. Rev. D {\bf 76}, 056007 (2007).
\bibitem{Diehl:2007jb} 
   M.~Diehl and D.Yu.~Ivanov, Eur. Phys. J. C {\bf 52}, 919 (2007).
\bibitem{Polyakov:1999gs} 
   M.V.~Polyakov and C.~Weiss, Phys. Rev. D {\bf 60}, 114017 (1999).
\bibitem{Guidal:2008ie} 
   M.~Guidal, Eur. Phys. J. A {\bf 37}, 319 (2008).
\bibitem{Kumericki:2009uq} 
   K.~Kumeri\v{c}ki and D.~Mueller, Nucl. Phys. B {\bf 841}, 1 (2010).
\bibitem{Goldstein:2010gu} 
   G.R.~Goldstein, J.~Osvaldo Gonzalez Hernandez and S.~Liuti, Phys. Rev. D {\bf 84} 034007 (2011).
\bibitem{Moutarde:2009fg} 
   H.~Moutarde, Phys. Rev D {\bf 79}, 094021 (2009).
\bibitem{Kumericki:2011rz} 
   K.~Kumeri\v{c}ki, D.~Müller and A.~Schäfer, JHEP {\bf 1107}, 073 (2011).
\bibitem{MunozCamacho:2006hx} 
   C.~Mu{\~n}oz~Camacho \textit{et al.}, Phys. Rev. Lett. {\bf 97}, 262002 (2006).
\bibitem{Goloskokov:2009ia} 
   S.~Goloskokov and P.~Kroll, Eur. Phys. J. C.  {\bf 65}, 137 (2010).
\bibitem{Boer:2011fh} 
   D.~Boer \textit{et al.}, Report of the joint BNL/INT/JLab program on the science case for an Electron-Ion Collider, Sept.~13 to Nov.~19, 2010, Institute for Nuclear Theory, Seattle.
\end{thebibliography}
\end{document}